# UFFENBACH'S "ZEITWEISER" PUBLISHED 1598

REINHARD FOLK, Institute for Theoretical Physics, University Linz

The time represented by the shadow of the sun on a sundial is a local time. Each place on earth (except on the same longitude) has its specific time. Mechanical clocks, synchronized to the local time, made this local time transportable. This enabled one to localize its position on earth relative to one, whose time was carried along, by comparing the local times of the two places. However mechanical clocks were not robust enough to keep their time when moved. Only by the development of Harrison's clock in the second half of the 18$^{th}$ century the clocks became robust enough to keep time with the necessary precision [1]. At the beginning of the 19$^{th}$ century, in the discussion of the electrodynamics of moving objects, local time (of course in another sense) returned [2, 3]. It turned out that time (and the time of mechanical clocks) is not absolute but depends on the movement of the clock and even the local gravitational field. We nowadays recognize that time is relative and we have to take this into account, when the GPS system is used to localize a position on earth. Thus a quite similar problem in localization existed in the 20$^{th}$ century as at the end of the 16$^{th}$ century, when navigators tried to find out their position on sea.

The gnomonic projection, used for making sundials, was understood already in ancient times but it was not used for mapping the sphere of earth on a plane. There are rare examples where this projection has been used for celestial maps. Johannes Kepler in 1606 used it for presenting his "stella nova" and later Christof Grienberger and Orazio Grassi for star charts [4]. There is only one geographical example known from 1610 by Franz Ritter [5]. In fact it was first published together with several horizontal sundials in his "Speculum solis" 1607 [6,7]. All known examples in the cartographic literature date after 1600.

However there can be found an earlier example, which combines time and space in the form of a sundial and a map, in the small book of Philipp Uffenbach "Bericht vnd Erklärung Zweyer beygelegten künstlichen Kupfferstücken / oder Zeitweiser der Sonnen / vber die gantze Welt" published in Frankfurt/Main by Brachfeldt in the year 1598 [8]. To my knowledge this work has never been described before.

2006 William J. H. Andrewes built a longitudinal sundial showing time on a gnomonic map of the earth [9]. On that occasion Dava Sobel described it and stated: ''The map's meridians of longitude would serve as the sundial's hour lines, creating a union of time and space for that particular location – something no dialist or clockmaker had ever before achieved'' and ''That map and others by the mathematician Franz Ritter are the oldest known examples of a gnomonic projection'' ([10] and see also [11]). Uffenbach's earlier work from 1598 was also in the spirit to unify time and space or, as one can say, of heaven and earth (see Fig. 1). It was also in the context of the tradition of printed instruments like astrolabes and sundials [12]. Most prominent representatives of this tradition were Georg Hartman, Philip Apian, Johannes Schöner and Bartholomäus Sculterius as Ritter mentioned in his *Speculum Solis* [7]. It is the aim of this note to present Uffenbach's 'longitudinal' sundial.

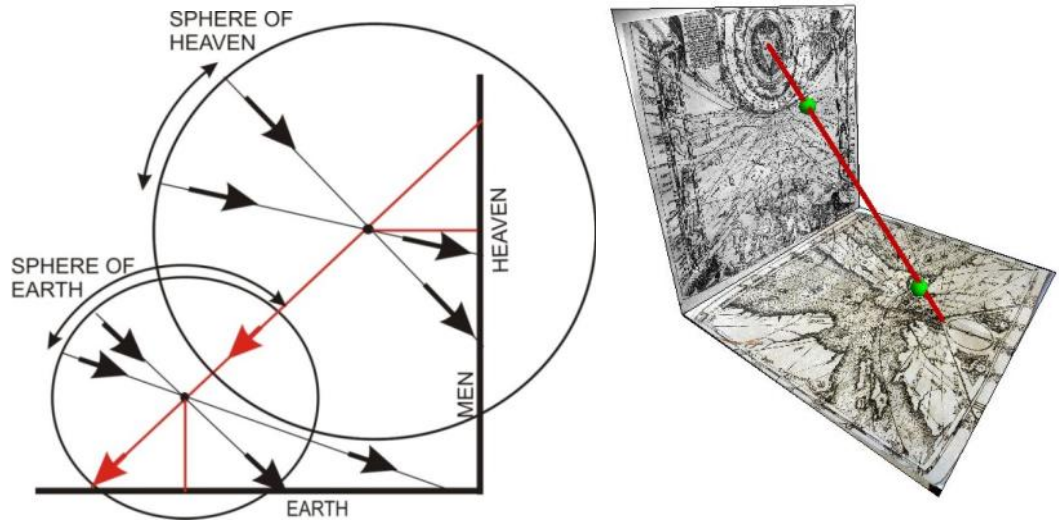

Figure 1: Left the two gnomonic projections of Uffenbach's diptych sundial seen on the right in an reconstruction. The gnomonic string connecting the intersection points of the German hour lines of the vertical and horizontal sundial contains two pearls indicating the two respective nodi of the sundial.

*The 'Zeitweiser' ('time pointer')*

Philipp Uffenbach, born 1566, originates from a famous family in Frankfurt/Main. His father Heinrich was a well-known block cutter and Philipp became also a painter, engraver and etcher. His late work as painter is strongly influenced by Mathias Grünewald and Albrecht Dürer. One should note that Frankfurt/Main was a protestant city since 1533, when the catholic masses where shut down. However in 1548 the Bartholomaeus church was opened for this purpose to the catholic minority.

Philipp Uffenbach took over the workshop of his father-in-law, when he married 1592 Margarete Hoffmann. In the year 1598, when he presented his "Zeitweiser" [6], he received the citizenship of Frankfurt/Main. He wanted to prove with the "Zeitweiser" his skill as artist, etcher and geometer. Moreover he wanted to assert his believe in God and to convey his view of the world. For that purpose he concentrated on "die vier gröste Wunderwerck GOTtes. Als Himmel/Erdt/die Zeit und der Mensch" (the four largest marvels of God: As heaven, earth, time and man).

In order to do so he fabricated a diptych sundial well known at that time but in addition to the usual time lines he added further information (see Fig. 1,2). These two important additions were (1) the twelve ascending signs of the zodiac and several bright stars shown on the gnomonic projections of the heaven and (2) a gnomonic world map (see Fig. 1,3). In his instruction for constructing the diptych sundial it is made clear that the two projections of heaven and earth are entangled by the gnomonic string and its shadow indicating local time. This is achieved by a simple advice when the two prints are glued to two vertical plates. One has to adjust two marks on the prints in such a way, that they should lie on each other, when folding the two plates of the sundial with the prints attached. This guaranties the correct angle (50 1/6 degrees for Frankfurt) of the gnomonic string independent of the distance of the two prints from the folding. Each nodus of the vertical and horizontal sundial is marked by a pearl on the string, which place is given in the instruction.

In the year 1619 Uffenbach published by himself the small work "De quadrature circuli mechanici... ", in Frankfurt and after he died in 1636, a second edition appeared 1636 in Nuremberg by Fürst. This is short tract on the problem of squaring the circle. It is the challenge of constructing a square with the same area as a given circle by using only a finite number of steps with compass and straightedge. But Uffenbach's tract is more concerned with practical matters than with theory. It deals, for example, with the layout of gear teeth and other design problems that would have been of concern to someone in Uffenbach's profession [13]. In the first chapter of this publication he cites some of his references books written by Albrecht Dürer, Simon Jacob and Andreas Helmreich. In Dürer's "Underweysung der Messung, mit dem Zirckel und Richtscheyt... " from 1525 as well as in Helmreich's "Rechenbuch" from 1595 in the fifth chapter, one finds instructions for construction of sundials. It is remarkable that in the section on the motion of the sun Helmreich [14] refers to Nicolaus Copernicus' "De revolutionibus orbium coelestium" (fourth book, chapter 19) for the astronomical data of the sun used.

*The heaven*

Usually one finds on a sundial a set of different types of hour lines also shown on Uffenbach's printed vertical sun dial. These are (see Fig. 2):

- German hour lines ("welche Ziffer in dem runden Zettel der Schatten verzeichnet"; 6 - 12 - 6, top circle around the word GOTT; in the graph below this circle very faint lines indicate the German hour lines also [15]), Babylonian (1 - 11) and Italian hour lines (13 - 24), both straight lines in Fig. 2 with small Arabic numbers.

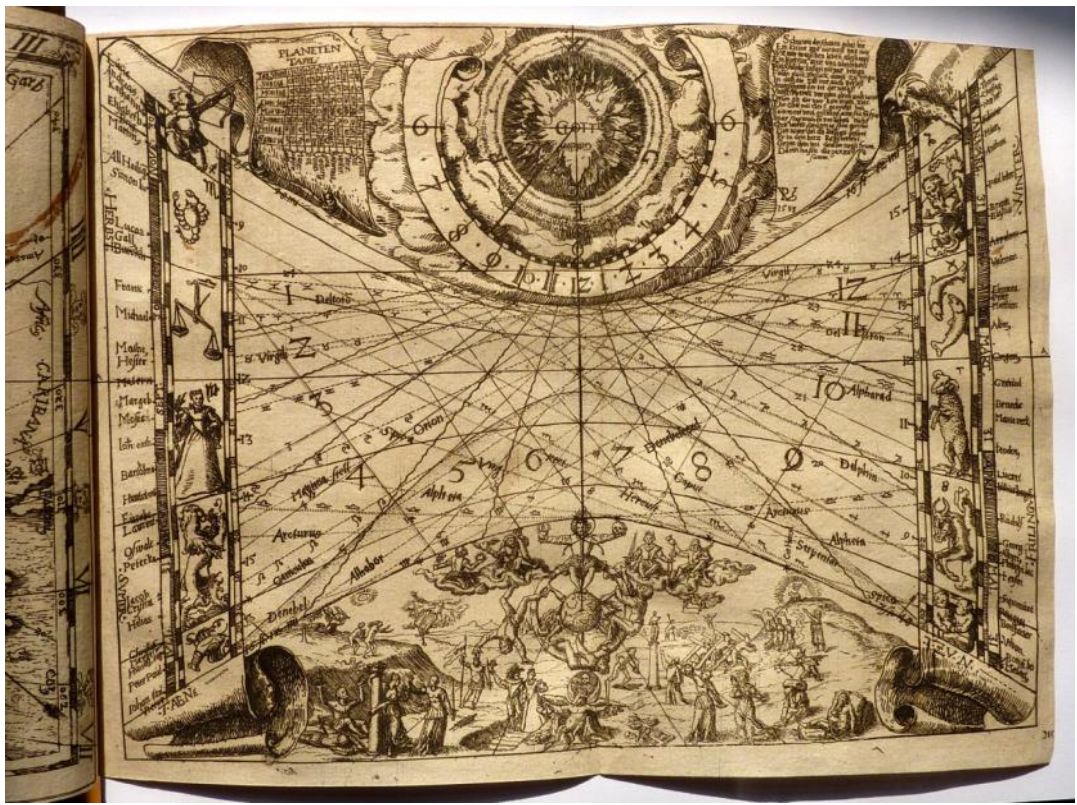

Figure 2: Vertical part of Uffenbach's diptych sun dial (courtesy monastery Chremsminster). For the lines and curves see text.

- Planetary hour lines (1 - 12, wiggled lines in Fig. 2 and large Arabic numbers between them) together with a planetary table, which indicates which planet rules over each hour (given by the large Arabic numbers on a certain day of the week.

The correctness of these lines can be easily checked by performing the gnomonic projection for the latitude of Frankfurt/Main which is noted by Uffenbach as 50 1/6 degrees. More difficult to check are the planetary hour lines since several different definitions can be used [16]. They can be taken either as (1) Apian lines, (2) temporary hour lines or (3) ecliptic planetary lines [17]. The last ones are not straight lines but complicated curves thus excluding case (3). In case (1) the hours are defined by rotating the horizon 15 degrees each hour along the axis to the south in the horizontal plane cutting twelve almost equal sections out of the diurnal arc. These hour lines should intersect in a single point [16, 17]. This is not the case for the planetary hour lines shown here. Thus it is concluded that Uffenbach shows temporary hour lines defined by the exact division of the diurnal arc into twelve parts. Although they are not the projection of great circles and therefore not straight lines they are in the region shown on the sundial 'almost' straight lines (see Fig. 12 in [16] and Ref. [18]).

As often on vertical sundials also Uffenbach's sundial shows the relation of the diurnal arcs to the signs of the zodiac and to the days of a calendar on the left and right side of the sundial. Since the equinox is found on Gregorius' day 12th of March (instead of 20./21.) a Julian calendar is shown. The protestants did not accept the Gregorian calendar reform of 1584 and the Julian calendar was in use in the year of publication of this work in Frankfurt.

Uffenbach also includes on the calendar the length of the day and/or the night at the diurnal arcs for the days, when a sign of zodiac is entered. One finds these data in a table of Helmreich's *Rechenbuch* [14]. If the shadow of the gnomonic pearl falls between two diurnal arcs one finds these values by interpolation. The longest day lasts 16 hours and 12 minutes for the latitude of 50 1/6 degrees.

Rarely found on sundials are the almucantar lines (altitude lines), but their construction can be found in the gnomonic literature of the time (see e.g. Andreas Schöner, "Gnomonice" from 562 [20]). Since the points of equal altitude lie on small circles, their projections are hyperbolas. An example of a sundial with almucantar lines can be found in Britain at the All Saints' Church in Isleworth [21]. On his sundial Uffenbach shows the almucantar line when the sun is just 45 degrees over the horizon (the shaded arch in Fig. 2). In such a case the shadow of a building has the same length as its height.

Also rarely found on sundials are

- the twelve ascendants indicated by their symbols together with
- thirteen brightest stars ("sampt der andern fürnemen Hauptstern") indicated by their names.

Of course these signs of zodiac and stars move on the sky according to the diurnal rotation of the earth, but is shown is the projection of their path through the year on the heavens sphere. This path is for each sign a great circle tangent to the small circles of the solstices. Their projections are straight lines touching the hyperbolas of Cancer and Capricorn. However one of these lines, that of Libra, is erroneously drawn with a kink and not parallel to the diurnal line of equinox. One cannot see the signs of zodiac or the stars during daylight but if the shadow of the pearl touches the respective line it is indicated that the sign of zodiac or star rises.

The projection lines of the stars are also straight lines since their annual path on heavens sphere is a great circle. The following stars are indicated (identifications are added in brackets [19]): Alhabor

(αCMa, Sirius), Alpharad (αHya, heart of the snake), Alpheta (αCrB, cameo), Arcturus (αBoo, guardian of the bear), Caniculus (αCma, Sirius, although 'caniculus' may be translated by 'little dog' it is the name used for Sirius [22]), Caput Serpentar (αOph, head of the serpent collector), Delphin (⬜Del), Deltoto, Deltoton (Triangle), Denebel, Denebeleced (βLeo, Denbola, the lion's tail), Maxima Stell Orion (βOri, Rigel, leftfoot), pect Herculi, Spica, Spica Virg (αVir, Virgo's ear of grain), Virgili, Virgilis (Pleiades). The Romans named the Plejades Vergiliae [23] and this was also used in some star charts (see for example Santbech's "Problemata #astronomicum..." from 1561 [24] or Honters "De Cosmographiae rudimentis ... " [22] from 1585).

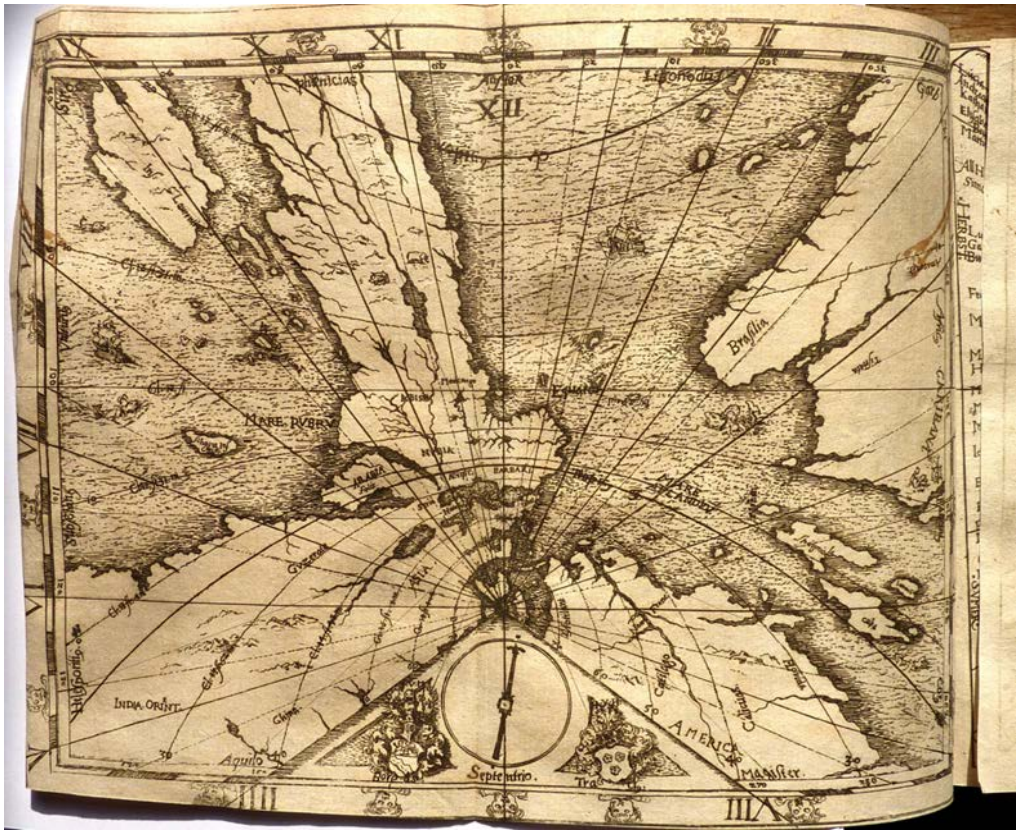

Figure 3: Uffenbach's gnomonic map and the horizontal part of his diptych sundial (courtesy monastery Chremsminster).

*The earth*

The horizontal part of the diptych sundial shows the local time in relation to the local time in Frankfurt as function of the location on a world map (see Fig. 3). The shadow of the string gnomon lies on one or between two longitudes seen in the projection as straight lines all crossing near the North Pole (allowing for a magnetic declination). The magnetic declination is shown on the map by a compass whose needle deviates from the true north south direction.

Orienting the horizontal sundial accordingly one can read off

- German hours (straight lines in Fig. 3 ending in the roman number of the hour, IIII - XII - VIII).

The meridian runs through Frankfurt/Main and the shadow shows the local time at Frankfurt and at all places on the same longitude. Local time on earth is due to the rotation of earth around its axis. Therefore a time distance of one hour corresponds to a distance of 15 degrees in longitude, or the other way round from the spatial difference of 15 degrees in the longitude one knows the corresponding local times differ by one hour from each other.

It was well known that one would be able to find out the distance of two longitudes if one knows the difference of their local times. But in order to know this time difference it should be possible to take along the local time of the reference longitude on one's journey. Only after the construction of Harrison's timekeepers in the 18th century the longitudinal problem was solved in a practicable manner.

In addition to the hour lines the map shows

- the projection of the longitudes for every ten degrees (alternating full and dashed lines in Fig. 3 from 0 - 160; 270 - 360)
- 16 Great circles through Frankfurt (every 22.5 degree; thin straight lines crossing in Frankfurt/Main) indicating the direction of corresponding winds named on the edges of the map.
- The maximal length of the day (denoted as C.L.) for each latitude (from the 70 degree north to the 20 degree south every 10 degrees). The curves of the projection are conic sections, usually hyperbolas but north to the polar cycle they are ellipses.
- The tropical circles (of Capricorn at 23.5 degrees south and Cancer at 23.5 degrees north) as the outermost latitudes where the sun can be in the zenith (at solstice in December and June, respectively).

It was common to show on maps and sundials the winds blowing from different directions. The winds were closely related to the sun, as the directions east and west are related to the rise and setting sun and north and south to the corresponding perpendicular directions of the meridian. The four directions were further divided and additional winds defined [25].

Starting from the south in clockwise rotation Uffenbach notes the following 16 winds: Auster S [Notus], Libonodus SSW [Libonotus, Austroafricus], *Garb* SW, Africus WSW [Lips, Liba], Favonius W [Zephyr], Corus WNW [Caurus, Argestes], *Magister* NW, Tractus NNW [Thrascias, Circius], Septentrio N [Aparctias], Boreas NNE, Aquilo NE, Helespontius ENE [Caecias], Subsolanus E [Apeliotes], Vulturius ESE [Volturnus], *Siroc* SE, phenicias SSE [Euronotus]. Either the Greek or the Latin names as found in [25] are given. The winds in italic have not been found in the usual compilations but they can be found on Italian or Portuguese mariner's compasses. Some of them are from Arabic root as Sciroc(co) (SO) from <al-Sharq = east> and Garb(ino) (SW) from <al-Gharb = west> - both were also translated by arising and setting [26]. These names are denoted on the world map of Honter's "De Cosmographiae rudimentis ... " [22]. Magister seems to be the French wind Magistral [Mistral].

The map shows all four continents known at that time: Europe, Asia, America and Africa. In addition several important geographical locations are named. In particular the islands in the ocean between Africa and South America were of great interest. Uffenbach's map shows four islands: Ins. S. Paul, S. Croce, Ascension (erroneously named S. Mathe) and S. Matheo (not named). They are all also shown on the map of Ortelius (1527 - 1598) from 1570 and the map of Hogenberg (1535 - 1590) from 1590 [26, 27]. In addition one can identify the unnamed Islands as Trinidad, Ascension, S. Maria, Yslas de Miriuaes (two islands I. of Martin) by comparing with these maps.

In the Caribbean Sea he charts Cuba, Iamaica, Spangiola (today Haiti and the Dominican Republic) and an island S. Johan which might be correctly named Borique (today Puerto Rico). The islands shown follow the map of the pacific in Ortelius Atlas from 1589 [29] or that of Hogenberg [30].

In the Arabic Sea (named Mare Rubrum) one finds the islands 'Schona Ins Christis' named on Ortelius' and Hogenberg's map Zacotora (today Socotra). On Waldseemüller's map it has its old name 'Scotira insula christana' together with a short description repeating a tale which the traveler Marco Polo disseminated. However the position of the island (between the longitude of 80 and 90 degrees) is erroneously shifted to the east by about 10 degrees compared to the maps mentioned. But surprisingly this shift corresponds to the position on the map of Waldseemüller from 1507, where the zero of the longitudes is shifted by 10 degrees.

Then south of the equator one finds the islands: Vasco de Acuna I. S.Spirito, IS. Liona and Ins. Laurentii. (today Mauritius). Again some other unnamed islands can be identified as S Francesco, Adarno.

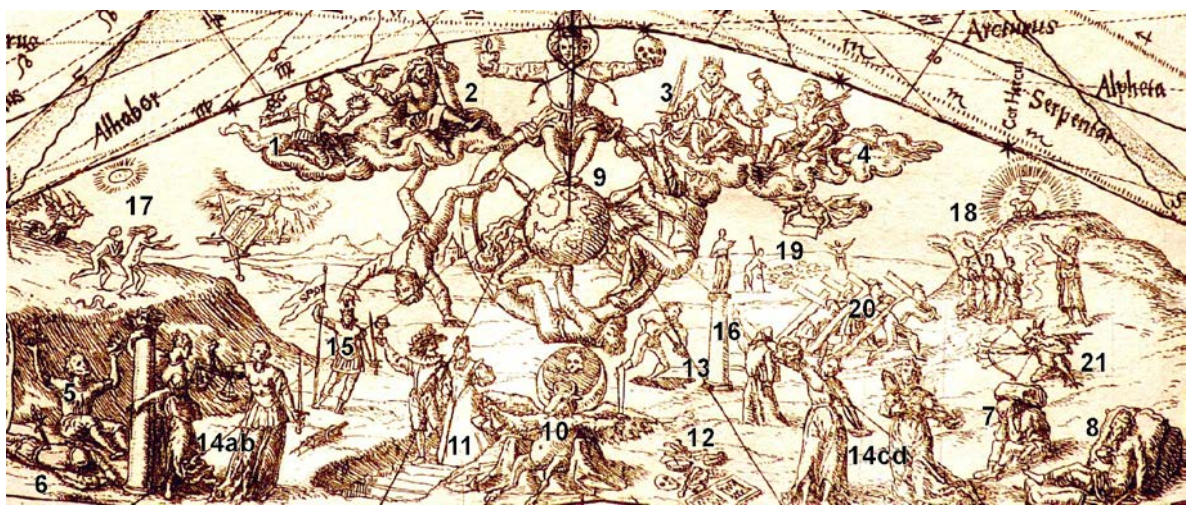

Figure 4: The picture below the vertical sundial (see Fig. 2) shows different scenes, which illustrate the course of life (for the different numbers see the text) (courtesy Monastery Chremsminster).

Whereas the further exploration of the heaven had to wait for invention of the telescope, the exploration of the earth by the circumnavigation proceeded continuously in Uffenbachs time. The painter Adam Elsheimer, Uffenbach's most famous pupil, etched 1597 a world map included in the work "Historicae relationis continuatio. Warhafftige Beschreibunge... durch Jacobum Francum" [i.e. Conrad Lautenbach]. Wallstatt [i.e. Frankfurt], 1598. The map illustrates the voyage of Cornelis Houtman, the first Dutch expedition to the East via the Cape of Good Hope in 1595-98 [31].

*The time men are given*

Like an instructions manual Uffenbach gave in the first part of his book advice how to construct the diptych sundial and explains what is shown on the vertical and horizontal sheet. However he made not only a printed instrument for daily use, but also wanted to show the creations of God: heaven (the sphere of stars and the signs of zodiac), earth (the world as it was known), and time given by the motion of the stellar sphere and the sun and men. For them the time of their lives is limited and therefore it should be used agreeable to God.

The combination of the prints and the religious text is reminiscent of Scipo's dream written by Cicero 1700 years ago. There Scipio Africanus in a dream looks from the milky way onto the cosmos and the earth. Under this impression he talks with his adoptive grandfather Cornelius Scipio Africanus (the elder) and his biological father Aemilius Paullus about the problem, how to arrange a virtuous life.

Cicero had some astronomical knowledge from his translation of Aratos *Phenomena*. From Uffenbach's presentation one can see the development in astronomy but even more in geography with respect to Cicero's knowledge but, as we know now, our understanding of the cosmos increased much more in the next century.

Regarding the life of men Uffenbach explains by referring to biblical scenes and allegories the fate a human being can take. These scenes are placed below the heavens spheres and symbolize the following items (see Fig. 4): 1: profane art 2: healthiness, strength 3: violence, wealth 4: peace, amity 5: loss of reputation 6: shortening of life 7: poverty 8: illness 9: wheel of life 10: monster of darkness 11: luxuriousness, gluttony 12: idleness, beguilement, lying 13: desperation 14: the four secular virtues a: courage, b: justice, c: temperance, d: prudence 15: pagan emperor 16: philosopher who adores his idol 17: expulsion from the Garden, Tables of the Law 18: Mount Zion, Lamb of God, John the Baptist 19: Annunciation to the shepherds 20: cross-carrying Christians 21: infernal archer

The central ''wheel of life'' connects the theological thoughts about life to the astronomical and astrological aspects of the diptych sundial. This is a common motive in medieval art reinterpreted in renaissance time [32]. Sometimes a ''memento mori'' can be found on ivory diptych sundials [33] reminding the user on the finiteness of life.

*Conclusion*

Uffenbach's work demonstrates on one hand his skill as craftsmen, geometer and engraver, but also as a religious person, whose faith rest on the bible. His work is based on the Aristotelian-Christian view of the world and has to be seen also in its time of religious conflicts. Apart from the religious content of the book and the graphic account several exceptional features are shown on the sundial like the lines of the ascending zodiacs and of the brightest stars. As an artist he presents us the oldest gnomonic world map known today in combination with a vertical sundial. The world map covers the region from Asia to America and from the North Pole down to Madagascar and Brasilia. Thus Uffenbach's sundial is a predecessor of the horizontal sundial published by Franz Ritter 1607 in Nuremberg.

*Acknowledgement:* I thank Ilse Fabian for making me clear the importance of this sundial and Harald Iro for comments on the manuscript. I thank the monastery Chremsminster for access to its library.